# QUERY OPTIMIZATION TECHNIQUES IN GRAPH DATABASES

Ali Ben Ammar[1]

[1]Higher Institute of Computer Science and Management
Kairouan University, Tunisia

## ABSTRACT :

*Graph databases (GDB) have recently been arisen to overcome the limits of traditional databases for storing and managing data with graph-like structure. Today, they represent a requirementfor many applications that manage graph-like data,like social networks.Most of the techniques, applied to optimize queries in graph databases, have been used in traditional databases, distribution systems,… or they are inspired from graph theory. However, their reuse in graph databases should take care of the main characteristics of graph databases, such as dynamic structure, highly interconnected data, and ability to efficiently access data relationships. In this paper, we survey the query optimization techniques in graph databases. In particular,we focus on the features they have introduced to improve querying graph-like data.*

## KEYWORDS

## 1 INTRODUCTION

The last decade is characterized by an explosion in applications managing data with graph-like structuresuch as social networks, telecommunication networks, linked webpages, ….These applications manage billions of interconnected data, which construct the graph structure. User queries in these applications are more interested on the relationships between data rather than on the nodes of the graph. As example of such queries we cite: For Facebook user, find its friends (neighbors) that have studied with him in the same college and that have worked in hospital; find the calls of a person in a telecommunication network; find the peoples buying a specific product or mining relationships between customers in a marketing study,….Theproliferationof these applications and the requirement to access complex data relationships have imposed the re-arise of graph databases (GDB) to be the most efficient systems forstoring and queryinggraph data. The attempts of implementing graph databases have been introduced more than 20 years ago[1], [2] but the ability of traditional database systems, especially relational databases and XML (eXtensible Markup Language), to manage small data graph, over the past time, has carried over thesuccess of these attempts. Today, traditional databases are enable to store large graphs and querying their complex relationships, which need complex joins of relational tables.The actual NoSQL (no structured query language) databases systems, like MapReduce, have demonstrated their efficiency to store large volume of historic data but this efficiency decreases when queryinghighly interconnected data with dynamic structure. Thus, there is a research tendency to graph databases(GDB)in order to develop standard tools for managing graph data. Researchers in this area exploit a large baseline of concepts and algorithms developed in graph theory over the last thirty years.

Graph database systems are particularly used to store and query labeled data graphs, which are sets of vertices and edges relating vertices. Vertices and edges have proprieties and each edge have a label describing the relationship between pair of vertices. For example a relationship between pair of Facebook users may be "friendship", "following" or "married". What distinguishes graph databases from traditional storage systems is that every vertex stores direct pointers to its adjacent elements. This specific feature, which allows querying neighborhood, constitutes an advantage to improve response time of user queries. User queries in graph theory are mainly pattern matching, data mining, and aggregation. We describe this queries with details insection 4. However, querying neighborhood alone is not sufficient to obtain satisfactory query response time. Many approaches, which benefit from the rich background on graph theory, have been recently developed to improve query response time in graph databases. In this paper, we addresstechniques used or shared by these approaches to optimize user queries. These techniques have been, previously, applied in other domains like databases, distribution systems, data stream, …. Particularly, we focus on the specificities of applying them in graph databases i.e. what are the specific reasons for their application in this research area and, given the underline data structured, how they have resolved the addressed issues. Our contribution is to studyrecent query optimization techniques in graph database. We summarize their key features, outline their current use cases and identify their limits. Recent studies in this area have addressed models, query languages and storage systems. To the best of our knowledge, we are the first that focus on the view of query optimization.

The rest of this paper is structured as follows. The next section discusses existing studies on graph database. Section 3 presents an overview about graph databases. It contains definition, design models and historic of graph databases. Section 4 presents the techniques of query optimization. We discuss these techniques in section 5 and in section6 we conclude.

## 2 RELATED WORKS

Graph databases are new storage systems that manage highly interconnected data. They propose typical solutions for recent applications in domains, dealing withgraph-like datastructure, such as social networks, telecommunications, biology, marketing, spatial analysis, criminal networks,…. The expected efficiency of graph databases has attracted the attention of database community and engaged researchersin the process ofdeveloping standard tools for design models and query languages, which represent the two main elements of any database project.Therefore, the early survey on graph databases is presented in [3]. It consists of a well-rounded survey of graph data models and their features. The paper has given multiple comparisons of graph data models with respect to data storing, data structure, query languages, and integrity constraints. Later, [4] has identify the current applications and implementations of graph databases.[5] provides a survey of many of the graph query languages, focusing on the core functionality provided in these languages such as subgraph matching, finding nodes connected by paths, comparing and returning paths, aggregation, node creation, and approximate matching and ranking. The authors have used the expressive power and the computational complexity as criteria to evaluate queries and then compare languages.[6]presents a performance introspection framework for graph databases, PIG, which provides both a toolset and methodology for understanding graph database performance. Then, authors have demonstrated the efficacy of this their framework by analyzing the popular Neo4j and DEX graph databases.  Graph databases models and languages are also recently surveyed in [7],[8]. In this paper, we address another view of graph databases, the query optimization. Recently, a few number of techniques have been proposed to optimize graph database queries. In this paper, we focus on these techniques and provide overviews of them, their features, their current use cases and their limits. To the best of our knowledge, this topic havenever been addressedbefore.

## 3 GRAPH DATABASES

### 3.1 Definition and examples of graph databases.

All the studied papers, in this work, agree that a graph database is a system to represent, storeand manage data, which are naturally interconnected and structured in a form of graph. The represented data are those contained in both nodes (vertices) and edges of the graph. According to[6], the term graph database embodies two main characteristics: (i) First, a graph database is a storage system whose native representation of data is in terms of objects (vertices) and inter-object relationships (edges);(ii)Second, a graph database supports single-object access via indexed lookup or iteration. That is, in addition to enabling whole-graph analysis, we require that a graph database be able to efficiently answer queries about the attributes and relationships of specific elements.

The authors of [9]provide a formal definition of graph database using graph terminology. They consider a graph databaseas a finite edge-labeled graph. i.e. Let $\sum$ be a finite alphabet, and V a countably infinite set of node ids,then a graph database over $\sum$ is a pair$G = (N, E)$, where $N$ is the set of nodes (afinite subset of $V$), and $E$ is the set of edges, i.e., $E \subseteq N \times \sum \times N$. That is, we view each edgeas a triple $(n, a, n')$, whose interpretation, of course, is an $a$-labeled edge from $n$ to $n'$.

The generalization of graph database instances is called schema or model and it is mandatory for the design and implementation of graph databases. The different models of graph databases are studied in [3]. The two main elements that make up all these models are nodes and relationships. The nodes represents entities and can hold any number of attributes. They can be tagged with labels representing their different roles in the studied domain. The relationships provide directed connections between two nodes.In most cases, relationships have quantitative properties, such as weights, costs, distances, ratings, time intervals, or strengths. Because of these properties, two nodes can share any number of relationships.

Example. The figure 1 presents schemas and instances of two graph databases. Figure 1.(a) represents an instance ofsome relationships in a social network like friendships and user-post relationships.  The generalization of this instance is described in figure 1.(c). Figure 1.(b) represents an instance of phone calls used to identify criminals   [a use case with Neo4j: http://neo4j.com/blog/use-phone-calls-identify-criminals/].The generalization of this instance is described in figure 1.(d).

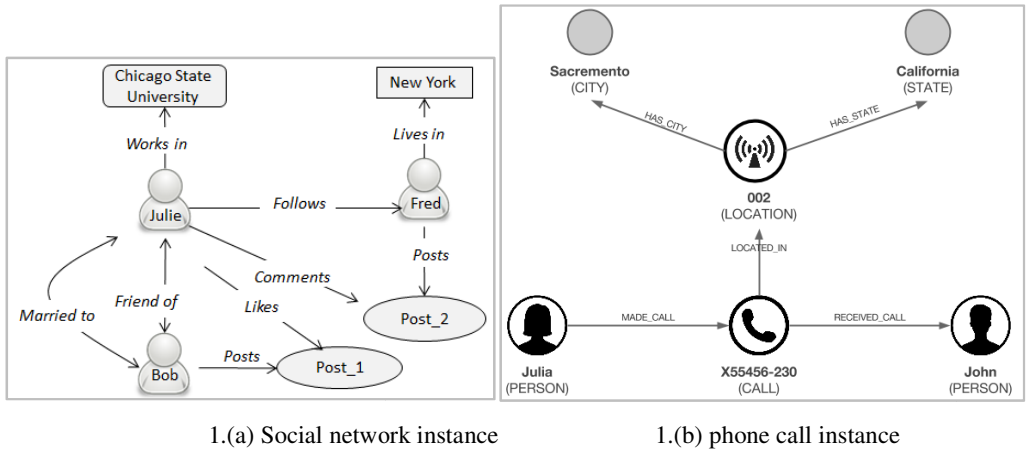

1.(a) Social network instance    1.(b) phone call instance

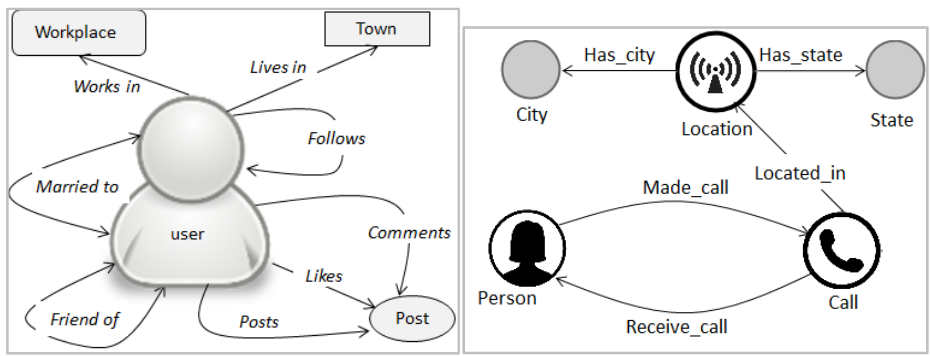

1.(c) GDB schema for social network    1.(d) GDB schema for phone calls.

Figure 1. Examples of graph databases (instances and schema).

Recently,[10]hasproposed an approach to derive a graph database schema from the classical entity-relation model. The main goal of this approach is the minimization of data access operations needed in graph traversals at query time. Intuitively, this can be achieved in two different ways: (i) by adding edges between nodes or (ii) by merging different nodes.

The explosion of social networks over the last few years has encouragedresearchers in GDB area to be more interested in such networks. Moreover, GDB are used to study and mine data of others domains like the phone calls in telecommunications, the protein interactions and neural pathways in biology, the relationships between people, themedia they use and the products they buy in marketing.

Graph nodes in certain applications may themselves comprise graphs like hypertext links. In this case we call the overall graph *hypergraph* and its model is more elaborated than the simple model described above.

International Journal of Database Management Systems ( IJDMS ) Vol.8, No.4, August 2016

## 3.2 Implementation

Recently, a limit number of models and systems are developed to implement graph databases such as Neo4j, DEX,….These tools are well studied and compared in[3], [11], [4], [6]. These studieshave demonstrated that Neo4j [http://neo4j.com/] dominates the others systems especially when querying graph databases.

Neo4j is a commercially supported open-source graph database. It is based on the data model of a directed multigraph with edge labels and optional node and edge properties. In Neo4j, both nodes and relationships can contain properties. Nodes are often used to represent entities, but depending on the domain relationships may be used for that purpose as well. A part from properties and relationships, nodes can also be labeled with zero or more labels. Figure 2represents the Neo4j model. [http://neo4j.com/developer/guide-data-modeling/].

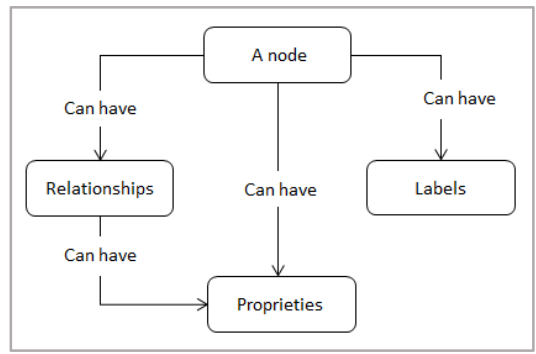

Figure 2.Neo4j graph database model.

The migration of databases from a relational model to a graph-based storage system (example Neo4j) respects a number of rules from which we cite here the main ones:
- Each entity table is represented by a label on nodes
- Each row in an entity table is a node
- Columns on those tables become node properties.

Further details about graph database systems and their performance are presented in [11], [6].

## 3.3 Historic and motivations

Storing data in a graph-like structure had attracted the attentionof database community in the first half of the nineties. After that, these attemptshad rapidly disappeared and this interest died with the uprising of XML and the Internet. Consequently, the database community moved toward semistructured data and people working on graph databases moved to particular applications like special data, web, and documents. Further details about the historic evolution of graph databases are provided in [3].However,along this period and since the eighties,the relational data model has dominated database management systems for more than thirty years.It has proved to be a powerful platform for business applications mainly for storing and retrievingdata.  It is characterized by its rigid schema and its well-known querying language SQL. The demand for alternatives to the relational model has grown with the rise of social media and the increase of massive and complex graph-like data introduced by the web. For example, Google, Facebook and Amazon have long been generating massive amounts of graph-like data using countless numbers of servers. Querying such data by a relational database system will need a complex SQL joins of multiple distributed tables. Thus, querying such data by traditional database management systems becomes impossible.Consequently, the limitations of traditional databasesto cover the requirements of current application domains, has ledto the development of new technologies, called NOSQL databases like Big Data and Graph Databases.

Recently, there is an interest to graph databases. This interest is due in part to the large amounts of graph data introduced by the Web like social media and in another part to the increase of applications that focus on the relationships between data more than on the data nodes. Moreover, graph databases represent the natural way for modelling graph-like data. Compared to the others NOSQL databases, graph databases are dedicate for managing higher level of data complexity. [4]has categorized the NOSQL databases as in the figure 3.

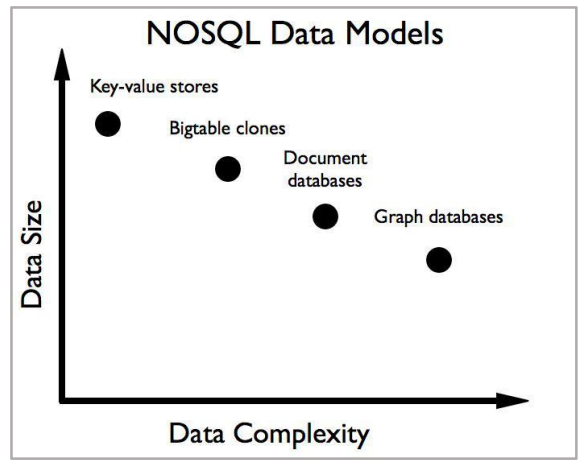

Figure 3.Categorization of NOSQL databases[4].

The main advantages or the powerof graph databases are presented in [8] as follow:

- Performance: The Graph database performance increases when dealing with connected data versus relational databases and NOSQL stores. In contrast to relational databases,Graph database systems provide direct access to relationships and perform well the join of multiple data sources.
- Flexibility: Graph database allows dynamic structure and schema i.e. we can add new kinds of relationships, new nodes, and new subgraphs to an existing structure without disturbing existing queries and application functionality.because of this flexibility, we don't have to model our domain in exhaustive detail ahead of time.
- Agility: Modern graph databases equip us to perform frictionless development and graceful systems maintenance. In particular, the schema-free nature of the graph data model empowers us to evolve an application in a controlled manner.

In addition, a graph database system is considered efficient if it supports selecting vertices by their proprieties [6]. For example, in a social network, we may select vertices that might represent people by their proprieties such as names, ages, etc.

International Journal of Database Management Systems ( IJDMS ) Vol.8, No.4, August 2016

## 4 QUERYING GRAPH DATABASES

Querying graph databases has started to be investigated some 25 years ago. Today, the proliferation of applications that manage high linked data has imposed the resurgence of the issue of how efficiently querying graph databases. Consequently, number of query languages and algorithms for optimizing query in graph databases have recently developed. In this section, we study the major techniques used to optimize queries in graph databases. We refer readers to [5]for further detailsabout query languages for graph databases.

The following basic rule: "Every graph node contains a direct pointer to its adjacent elements and no index lookups are necessary" make the apparent distinction between graph databases systems and the traditional DBMS. Therefore, all query languages for graph databases are based on this rule and the algorithms, studied for tens of years in graph theory, constitutes a solid root that underline them.

[5]compares the efficiency of some query languages for graph databases. They have used a list of queries as criteria of comparison. They classify the used queries into three approximate levels. The lower level contains database API operations, called micro-operations, such as "GetVertex/Edge" or "SetProperty". The intermediate level contains operations that aggregate API calls, such as "Get Neighbors" or "Ingest". These operations are calledgraph operations. The upper level corresponds to algorithms that express fundamental concepts in graph algorithms, but are often not part of a graph API, such as "FindShortest-Path", "Compute Clustering Coefficient".

### 4.1 Graph query terminology

- *Pattern* :

A pattern graph (or pattern query) $P = (V_P, P_E)$ specifies the structural and semantic requirements that a subgraph of G must satisfy in order to match the pattern $P$. The task is to find the set $M$ of subgraphs of $G$ that "*match*" the pattern $P$. A graph $G' = (V', E')$ is a subgraph of $G$ if and only if $V' \subseteq V$ and $E' \subseteq E$[12]. Here we are based on the formal definition of graph database in section 3.

The following example, provided in[13], presents a pattern query and the steps to execute it in the context of social network whose graph database schema may corresponds to the model in figure 1: For a person $x$, find all of his $friends$ who have worked in $Stanford\ Business\ School$ and who have a $friend$ from $South\ Africa$. For a query like this, we have to start from node $x$, visit all its neighbors to check which of them had worked in$Stanford\ business\ school$, and then for all those friends, visit their neighbors till we find a neighbor matching the predicate.

- *Graph matching:*

The goal of graph matching is to find occurrences of a specific pattern in a graph [12]. A match or a partial match is as a set of edge pairs. Each edge pair represents a mapping between an edge in a query graph and its corresponding edge in the data graph [14]. In simply form, a match is an isomorphic subgraph[14]. Some researchers propose semantic matching approaches that attempt to match graphs based on their meaning by taking into account vertex and edge types and attributes as well as graph structure.
Conjunctive queries, regular path queries and conjunctive regular path queries, which are presented in [5] and supported by most query languages in graph databases, are all forms of pattern matching. Conjunctive queries correspond the simple definition of pattern cited above.

Regular path queries consists of finding all node pairs (x,y) such that there is a path, from node x to node y, whosesequence of edge labels matches some pattern. This sequence of edges represents the researched regular path. Conjunctive regular path queries are the combination of conjunctive queries with the regular path ones.

- *Graph mining:*

Consists of finding a set of the most common or most "interesting" patterns in a graph [12]

- *Continuous query:*

Continuous queries are distinguished from ad-hoc query processing by their high selectivity, i.e. they look for unique events. Their aim is to detect newer updates of interest and notify a listener as soon as the query is matched. In this sense they are trigger-oriented[13].

- *Aggregate queries or ego-centric queries:*

Aggregate queries are used to calculate some proprieties of the graph. For some applications, they store theiroutputson nodes, which are part of their inputs;however, for others they create new nodes. Recently, aggregate queries have been used in [15], [16]to aggregate events coming from sources nodes, called producers, and stored in graph nodes, called consumers. In this context, an aggregate query is a producer of a single stream of events on a given topic from multiple sources.

## 4.2 Optimization techniques.

The major part of query optimization techniques in graph databases, and particularly those studied in this section, are previously applied in traditional databases. The interest of studying them here is to focus on the specificities of applying these techniques in graph databases and to describe the new ones of them like graph sketching. In addition, we discuss these techniques in the next section.

### 4.2.1 Data Distribution

This technique consists of partitioning data over several servers when it is infeasible to query and store them on a single site. The aim is to improve query response time and to minimize the data storage cost.Distribution has been extensively studied in traditional databases and often known as data replication [17], [18], and then applied in MapReduce framework forstoring and managing massive data[19], [20].

Graph databases are characterized by real-time data ingest, which make their schema dynamic, and real-time querying. In addition, most queries in graph databases traverse edges to fetch neighbor's information. For these reasons and because of the large size of graphs, theproblem of optimally partitioning a graph into equal-sized or equal-weighted partitions while minimizing the edges cut is NP-Hard. Consequently, graph partition techniques should found a good compromise between reducing server communication cost, which is defined in most cases by the number of edge-cuts, andreducing server-processing loads.

To do this, the studied works have applied different approaches thatdiverge in:

- ▪ The number of steps of the partition algorithm.

- The criteria used to partition
- Update of the partition
- Storage of auxiliary data
- Replication

To construct partitions, most of the studied approaches are based on the community structure to group the graph nodes. Community contains nodes and their neighbors.The aim is to reduce the edge-cut load. This concept is strictly applied in [21], which imposes that all the neighbors of a vertex $v$ should be replicated locally in the same partition. Although this grantees the local sematic, reduce edge-cut load but it increases the update load because of the high number of replicated data. This constraint is mitigate in [13] where the authors have introduced a novel parameter called *fairness requirement* which is characterized by a threshold $\tau \leq 1$ that means for each vertex we require that at least a $\tau$ fraction of its neighbors be present locally in the same partition.This fairness criterion is verified at the end of each repartitioning iteration. The approach proposed in [13]begin with an initial partitioning based on a hash function. Then it calibrates the result partitions by grouping together nodes having the same access pattern. Access pattern means whether the read frequency of a vertex $v$is low or not. Moreover, the access pattern tells us about the rate of read queries that ask both $v$ and its neighbors. The access pattern is used by the system to assign, for each partition, an update policy i.e. either up-to-date or lazy. The aim is to reduce the partitioning cost.

The use ofnatural community structure topartition dynamic graph was also applied in[22], [23]. The partitioning approach presented in[22]start with initial partitions created by a hash function. Then the incremental partitioning is done by migration of vertices from source partitions to receiver ones. At each iteration, every vertex $v$should decide to migrateto a partition containingsome of its neighbors or to remain in the current partition. But, in order to maintain a balanced load of servers, $v$ must stay in the current partition if this later contains some of $v$ neighbors. In [23], the authors propose a multi-level label propagation (MLP) method for graph partitioning. Label propagation (LP), which was originally proposed for community detection in social networks, is an iterative process. After assigning a unique label id to each vertex, the process update the vertex label iteratively. In each iteration, a vertex takes the label that is prevalent in its neighborhood as its own label. The process terminates when labels no longer change. As a result, vertices that have the same label belong to the same partition.

In a recent work [24], a dynamic partitioning algorithm called *lightweight repartitioner* was proposed. This approach overcomes the overload of incremental partitioning, characterizing the previous approaches, and the limit of load imbalance, presented in some approaches. To perform the incremental partitioning, the algorithm relies on a small amount of knowledge on the graph structure called auxiliary data. These data represent the accumulated weight of vertices in each partition and the number of neighbors of each hosted vertex in each partition. To guarantee a balanced load between partitions, the proposed algorithm assign weights to vertex and partitions based on the heuristic proposed in [25]. This heuristic consists of partitioning graph into $\alpha$ partitions where the weight of a partition $P$ is considered as the total weight of its vertices. According to this heuristic, a partitioning solution is considered valid if the weight of each partition is at most a factor $\gamma \leq 1$ away from the average weight of partitions. In other words, each partition $P$ of a graph $G$, should respect the following constraint: $w(P) \leq \gamma \times \sum_{v \in V(G)} w(v)/\alpha$ where $w(P)$ and $w(v)$ denote the weight of a partition $P$ and vertex $v$, respectively. The parameter $\gamma$, which is in range [1,2] is called the imbalance load factor and defines how imbalanced the partitions are allowed to be. When $\gamma = 1$ then partitions are required to be completely balanced (all have the same aggregate weights) and when $\gamma = 2$ then this allows the weight of one partition to be up to twice the average weight of all partitions.To calculate vertex weights, the algorithm proposed in [24]was consider the read requests to a vertex $v$as its weight.

Thus, graph-partitioning approaches should assign vertices to sites based on how much this assignment will improve query response time. So, to expect this query time, the most of them have consider the number of edge-cuts between partitions to simulate the query workload. However, the contribution of the approach, presented in [26], is to take care of the hardware features when expecting workload. In this contribution, the main factors to be considered are the volume of transferred data, the network heterogeneity and the services provided by the underline hardware architecture like the shared resource contention in multicore systems.

As we have seen in the above summary of the recent approaches of dynamic graph partitioning, replication and incremental partitioning have been applied in the most of these approaches. The role of replication is to reduce partition load and minimize the number of edge-cuts. The role of incremental partitioning is to continually maintain balanced load between partition and to improve the overall query response time. The advantage of Hermes [24], a system developed as an extension to the open source Neo4j to support workload partitioning, is the fast and efficient update of the partitionscheme. Another advantage of this system is the incremental integration of news queries. It has joined the partitioning solution proposed in [13]. This feature represents the limit of the solution proposed in [26]. The portioning solution proposed in [23]does not allow replication, which represents its major limit.

### 4.2.2 Query decomposition and incremental processing

There is no much work in query decomposition and incremental processing in dynamic graph area. Moreover, incremental processing and query decomposition are mainly devoted for graph pattern matching [14], [27], [28]. Incremental processing is used as a technique to avoid running complex queries from scratchon large graphs whenever they launched or the data graphs is updated.The aim is to minimize unnecessary recompilation and improve response time. With each new update on the data graph, the process of incremental processing checks to see if there are new results to the designed queries. In the context of continuous queries, it notifies the listener when there are new results forits queries. Otherwise,it plays the role of pre-processor for the queries.The main idea of query decomposition technique is to split the original graph query into subgraph ones. After finding matches for these subqueries, the process return up to assembly the sub-results. Query decomposition and incremental processing are complementary techniques. They are used separately in [27],[28] and combined in [14]. [27]presents incremental algorithms for graph pattern matching and discusses their complexities.[28]implements a decomposition based algorithm for searching a large static graph in a distributed environment.It splits any matching query into a set of subquery graphs that can be efficiently processed via in-memory graph exploration. The decomposition is well performed in order to avoid expensive join operations when merging sub-results. Another advantage of this approach is that graph indexes are unused which minimizes server loads. In contrast to these two works, [14] presents a complete solution for optimizing queries by the use of decomposition and the incremental processing. Another feature that distinguishes the later approach is the focus on continuous queries in dynamic graph. In such situation (dynamic environment or streaming in general) the use of indexation to accelerate queries will be a costly process since index should be update or recomputed after each graph update. Also, the periodic preprocess of queries will be wasteful especially when graph update are less frequent. The solution, proposed in [14], consists of an incremental processing of queries. It consists of finding the set of matches that result from updating the graph with a set of new updates, especially edge and vertex ingests.

The idea of the algorithm proposed in [14] is to decompose a graph query $G_q$ into subgraphs. Then we construct a binary tree called SJ-Tree so that each node of this treewill represent a subgraph of $G_q$ and that this subgraph, except those of the leaf nodes, corresponds to the combination of the two subgraphs of their children nodes. So, the root node will corresponds to $G_q$. The nodes of this tree stores references to all their correspondent matches retrieved on the dynamic graph. When a new edge is added to the data graph we check to see if there is a match to be inserted to leaf nodes. When a new match is inserted we check to see if it can be combined with any match of the sibling node. Each successful combination leads to the insertion of a larger match at the parent node. This process is repeated recursively and it terminates either by a complete matching in the root node or by a combination failure at any iteration of the algorithm. A problem that faced this solution when decomposing graph query is that if we choose a small subgraphs we undertake an overload of searching frequent matches. However, if we choose larger subgraphs we undertake an overload of searching complex matches. To overcome this problem, the authors of [14] have extended the original algorithm to consider lazy search of matches. The idea is that, when a new edge is added to the graph data, we firstly searching the node (the subgraph) with less probability (with high selectivity) to findmatches to it. If there is no match for that node, we stop searching for the rest of subgraphs since a part of the query is not satisfied. This algorithm performs well the decomposition but it suffer from the cost of managing auxiliary data.

### 4.2.3 Graph sketching

This technique consists of summarizing the behavior of large dynamic graphs like social network and web graphs. Dynamic graphs are characterized by a stream of edge insertions and deletions. Queries that are interested in controlling graph behavior will be affected by this rapid change on the data structure. Moreover, their repetitive execution on the data graph will be costly. So,the aim from creating a sketch is to compute the proprieties of this evolving graph without dealing (storing, updating) with the entire graph and then to enable effective query processing. The computed proprieties may be numbers of nodes and edges, distances between edges, frequency of edges, …In other words, graph sketch are simply a synopsis data structure that allows good approximations of the relevant properties of the data set[29].Consequently, graph sketch are limited to serve aggregate-like queries or analytical requests. [30]proposes a graph sketch method to estimate and optimize the response to basic queries on graph streams. The typical queries considered in [30] are edge query and aggregate-subgraph query. Edge query allows to estimate the frequency of particular edges in a graph stream. For example, in social network context, the estimation of the communication frequency between two specific friends is an edge query. However, aggregate subgraph query used to determinate the aggregate frequency behavior of constituent edges of a subgraph. For example, in social network context, the estimation of the overall communication frequencies within a community is an aggregate subgraph query.[29]proposessketch-based algorithms for estimating shortest path distances, frequency of various subgraphs and sparsification. A recent sketch-based approach, presented in [31], has used sketches in graph clustering. Graph clustering allows identifying communities or dense subgraphs in the graph and it is important to network structure inference, anomaly detection, and data mining.

### 4.2.4 Pre-processing aggregate queries.

Aggregate queries are characterized by a cost joins and a complex calculus. In traditional databases, materialization has proven to be a good technique for optimizing response time of aggregate queries. It allows pre-computing and storing their results, called views, to avoid re-computing them whenever they are asked. However, this technique may be not efficient in dynamic environments like graph databases and continuous stream. In dynamic environments, materialized views need to be updated continually, which increases their maintenance cost and decreases their efficiency.The previous technique, i.e. sketching and incremental processing may be used to optimize aggregate queries. To the best of our knowledge, there is no complete approach that optimizes aggregate queries regardless of its form or pre-process them. But, the ego-centric aggregate queries, which is a special case of aggregate queries, were addressed in [16], [15]. Ego-centric queries allow to a graph node, called consumer, to aggregate events from others nodes, called producers. For example, in social network, usersare often interested in events happening in the network but also physically close to them. So, theyare interested in an aggregate over the current state or the recent history of their neighborhood's nodes. Both [16] and [15] have addressed the issue of when the events should be transferred from the produces to the consumer. The two possible ways are: either at query time or materialized in advance on the consumer. The former way consists of traversing the producers at each read of the consumer. In [16], this way is called *pull* task and it corresponds to an on-demand update of aggregate data. However, the later one consists of pre-computing aggregate query answer at each new write in the producers. In [16], this way is called *push* task and it corresponds to online update. [15]proposesto retrieve events from high-rate producers at query time and materialize, in aggregation nodes, events that come from high-rate producers. [16] proposes detailed solution that beginsby constructing an aggregation overlay graph and then makes a decision for each node of this graph whether to aggregate events on it (push decision) or not (pull decision). The aggregation overlay graph is constructed to encode the computations to be performed when an update or a query is received. The main advantage of the aggregation overlay graph is that it allows sharing partial aggregates across different ego-centric queries. The decisions of materializing events on nodes are made based on the cost of push and pull tasks. As we have seen, what distinguishes [16] from [15] is the response to the issue of where do we store materialized data? For this reason, the solution of [16] has integrated intermediate aggregation nodes.

## 5 DISCUSSION

As in traditional databases, queries in graph databases cover the transactional operations, known in traditional databases, to add, delete, update and find entities. The underline data structure of GDB allows running efficiently queries with complex joins, called patterns. Pattern matching is the most query type required on graph database applications. They are considered as analytical queries that allow online finding semantic relationships of baseline data in order to discover knowledges from the data graph.Pattern matching may be processed in continuous way, where for each new events we check if a new match of the pattern is reached. Therefore, compared to data warehouse and big data, historic data have less importance in analytical queries in graph databases.

Query optimization techniques, which are inspired from distribution systems, graph theory, traditional databases and streaming domains,handle problems arisen from three characteristics of graph databases: large volume of stored data, high-interconnected data and dynamic schema of graph databases. Contrary to traditional databases, algorithms for optimizing graph queries have arisen before the born of graph database systems. They have been widely discussed in graph theory tens of years ago. They constitute a baseline for attempts of optimizing queries and developing query languages for graph databases. These attempts are recent and with limited number and until now there is no standard query language for graph databases.In this paper, we have studied the query optimization techniques, which are recently developed on graph databases:

Structure distribution, query decomposition, incremental processing, sketching and aggregations.These techniques may be complementary, i.e. any combination of them is possible and may lead to good results. However, they suffer from constraints on the types of queries they treated. For example, aggregations are limited to ego-centric queries, incremental processing are limited to pattern matching,…. So, there is no technique developed to be used for all graph query types. In addition to that and in most cases, auxiliary data represents a limit of these techniques, which manage lot of such data to perform their tasks.

## 6 CONCLUSION

Graph databases represent an important requirement for applications that manage graph-like data. Recently, these applications have been proliferated especially on the web, like social networks. The concepts and algorithms of graph database area constitute an overlap between traditional domains such as database and graph theory. The database community benefit from this background to develop systems managing graph databases. In this paper, we have presented an overview about graph databases. In particular, we have focused on the optimization techniques used to improve query response time. Most of these techniques have been applied in traditional databases and distributed systems such as query decomposition and pre-processing. We have identified the specificities of their current use in graph database and their limits. We conclude that the current use of all optimization techniques, studied in this paper,is limited to specific queries, i.e. there is no general technique to optimize any type of query in graph database. Moreover, there is no standard language to query graph database. Modeling database schema and decomposing queries are the most topics addressed by researchers.To summarize, we say that this area is in its beginning and much work is needed to satisfy the requirement of the recent applications and to support their proliferation, especially, on the web.